\documentstyle[epsfig]{elsart}

\def\be{\begin{equation}}
\def\ee{\end{equation}}
\def\bea{\begin{eqnarray}}
\def\eea{\end{eqnarray}}
\def\bk{{\bf k}}

\def\dst{\displaystyle\strut}
\def\ov{\over\displaystyle\strut}

\def\t0{\tau_0}

\def\bea{\begin{eqnarray}}
\def\eea{\end{eqnarray}}

\def\bk{{\bf{k}}}
\def\bK{{\bf{K}}}
\def\bK{{\bf{K}}}

\def\dst{\displaystyle\phantom{|}}
\def\ov{\over\dst}

\def\nb{{\overline{n}}}
\def\etab{{\overline{\eta}}}

\def\t{{\tau}}

\def\ben{\begin{eqnarray}}
\def\enn{\end{eqnarray}}
\def\bea{\begin{eqnarray}}
\def\eea{\end{eqnarray}}
\def\be{\begin{equation}}
\def\ee{\end{equation}}
\def\ov{\over\displaystyle\strut}
\def\dst{\displaystyle\phantom{|}}

\def\BL{{Buda}{-}{Lund}~} 

\begin{document}
\begin{frontmatter}
\title{\bf Model independent shape analysis of correlations\\
	in 1, 2 or 3 dimensions}

\author[KFKI]{T. Cs\"org\H o\thanksref{tamas}}
\author[KFKI]{and S. Hegyi\thanksref{sandor}}
\address[KFKI]{KFKI RMKI, H-1525 Budapest 114, POB 49, Hungary}

\thanks[tamas]{Email: csorgo@sunserv.kfki.hu}
\thanks[sandor]{\phantom{Email:} hegyi@rmki.kfki.hu}

\date{\today}

\begin{abstract}
A generic, model-independent method for the analysis of the
two-particle short-range correlations is presented, that can be
utilized to describe e.g. Bose-Einstein (HBT or GGLP),
statistical, dynamical or other short-range correlation functions. 
The method is based on a data-motivated choice for
the zero-th order approximation for the shape of the correlation
function, and on a systematic determination of the 
correction terms with the help of complete orthonormal
sets of functions. The Edgeworth expansion is obtained for 
approximately Gaussian, the Laguerre expansion for 
approximately exponential correlation functions. 
Multi-dimensional expansions are also introduced and discussed.
\end{abstract}
\end{frontmatter}

\section{Introduction}
Can one {\it model-independently}
characterize the shape of two-particle correlation functions?
In the present discussion,
we address such model-independent characterizations
of short-range correlations
on the level of statistical analysis. 
We do not make any theoretical assumptions,
neither on  the thermal or non-thermal nature of 
the particle emitting source,
nor on the negligibility of Coulomb and other final state interactions,
nor on the presence or the negligibility of a coherent component in the
source, nor on the negligibility of higher order quantum statistical
symmetrization effects, nor on the negligibility of 
dynamical effects (e.g. fractal structure of gluon-jets)
on the short-range part of the correlation function. 
We simply propose an expansion technique based on complete orthonormal
sets of functions and an experimental
choice for the zero-th order approximation to the BECF.

Although we do not assume here any theoretical 
knowledge about the short-range part of the two-particle
correlation function, we assume the following

\underline{\it experimental properties:}

{\it i) } 
The correlation function tends to a constant
for large values of the relative momentum $Q$.

{\it ii)}
The correlation function has a non-trivial structure at
a certain value of its argument. 

The location of the non-trivial structure in the correlation
function is assumed for simplicity
to be close to $Q = 0$.

The properties {\it i)} and {\it ii)}
are well satisfied by e.g. the conventionally used two-particle Bose-Einstein
correlation functions (BECF-s).
For an introduction to and a review of the 
recently determined non-ideal features
(e.g. non-Gaussian shapes in multi-dimensional 
Bose-Einstein correlation studies),
we recommend ref.~\cite{kittel}.

One of the parameters of the 
fit functions is the intercept parameter $\lambda_*$, that carries
important information
~\cite{Csorgo:1999vd,Vance:1998wd} on the possible restoration
of $U_A(1)$ symmetry in hot and dense matter, when the core-halo
picture can be applied~\cite{chalo}.
However, the intercept parameter $\lambda_*$ is known to be particularly
sensitive to the chosen form of the parameterization of the correlation 
function. For an exponential fitting function, the extrapolated intercept
is typically higher than that of the Gaussian fit, sometimes even
by a factor of two. However, both parameters have only a {\it few percent}
error from the minimization procedure~\cite{edge,edge-cst}.

It is well known~\cite{kittel}, 
that certain 1 or 2 dimensional correlation functions
can be  fitted with both exponential and Gaussian forms. In case of the 
NA22~\cite{na22} and the UA1 data~\cite{ua1}, 
the second-order correlation function
exhibits a stronger than Gaussian or exponential rise at low values of 
the invariant relative momentum $Q_I$. 
Recently, even singular parameterizations were shown to represent well the
precision data of the NA22 and the UA1 collaborations~\cite{na22,ua1}.
In this case, a formal interpolation of the fitting form to zero relative
momentum would yield infinite value for the intercept.
Furthermore, the intercept parameter $\lambda_*$ of the core-halo picture
is rather difficult to measure, as various non-ideal effects due to
detector resolution, particle mis-identification, resonance decays,
details of the Coulomb and strong final state interactions etc may
influence this parameter of the fit. One should also mention,
that if all of these difficulties are properly handled, the 
intercept parameter $\lambda_*$ for like-sign charged bosons is
(usually) not larger, than unity as a consequence of quantum statistics
for chaotic sources, even with a possible admixture of a coherent
component. However, attractive final state interactions,
fractal branching processes of gluon jets, or the
appearance of one-mode or two-mode squeezed states~\cite{squeezed,acg} 
in the particle emitting source might provide arbitrarily
large values for the intercept parameter.

Let us emphasize, that our new method is based {\it only} on the
properties ${\it i})$ and ${\it ii)}$ of the correlation function.
Thus, the method is {\it really} model-independent, and it can be
applied not only to Bose-Einstein correlation functions
but to any other experimentally determined function,
that satisfies properties ${\it i})$ and ${\it ii})$.

\section{Shape expansions using 
complete orthonormal sets of functions \label{s:gene}}
Let us define the two-particle correlation function as the
ratio of the two-particle invariant momentum distribution to the 
product of the single-particle invariant momentum distributions~\cite{misko}:
\be
C_2(\bk_1,\bk_2) = \frac{N_2(\bk_1,\bk_2)}{N_1(\bk_1)\, N_1(\bk_2)},
	\label{e:cdef}
\ee
where $\bk$ stands for the three-momentum of the detected particle
of mass $m$ and energy $E = \sqrt{m^2 + \bk^2}$.
For simplicity, we assume that at large relative momenta,
the two-particle correlation function tends to 1;
i.e. we normalize $C_2$ in such a manner that the possibly
existing long-range correlations are removed by the normalization,
which can be achieved by introducing an appropriate
normalization coefficient into eq.~(\ref{e:cdef}).
The two-particle correlator is introduced as
\be
R_2(\bk_1,\bk_2) = C_2(\bk_1,\bk_2) -1.
\ee
We assume that $R_2(\bk_1,\bk_2)$ is 
experimentally determined and it satisfies
properties {\it i)} and {\it ii)}.
Further, we assume that the measured
$R_2(\bk_1,\bk_2)$ is similar to a function $w(t)$,
where $t$ denotes certain experimentally defined
dimensionless combination(s)
of relative momentum variables, in 1, 2 or 3 dimensions. For
example, the experimental data are frequently represented
in one dimension in terms of $t \propto Q_I = \sqrt{ - (p_1 - p_2)^2}$,
or in three dimensions in terms of $t \propto (Q_0, Q_z,  Q_t) =
(E_1 - E_2, \bk_{1,z} - \bk_{2,z}, 
\sqrt{
(\bk_{1,x} - \bk_{2,x})^2+
(\bk_{1,y} - \bk_{2,y})^2} )$. We define $t$ to be a 
dimensionless variable, i.e. the relative momentum multiplied with
certain spatial scale, e.g. $t = Q_I R_I$. 

The function $w(t)$ can be considered simultaneously as a zero-th
order approximation to the function $R_2(t)$, as well as a certain
weight function or measure, that is related to an abstract Hilbert
space and a complete orthonormal set of functions, $h_n(t)$.
This property of the $w(t)$ function will be the basis of the new
expansion techique, and the expansion will consist of finding the
coefficients of the $h_n(t)$ functions.
For the application of the results to expand the real-valued
two-particle correlation functions $C_2(\bk_1,\bk_2)$,
we demand that all the $h_n(t)$ functions are real-valued
functions of the variable(s)  $t$.
We discuss those situations only, when such a complete orthonormal
set of functions exists with respect to the measure $d\mu(t) = dt\, w(t) $:

\bea
\int dt w(t) h_n(t) h_m(t) & = & \delta_{n,m},\\
f(t) & = & \sum_{n = 0}^{\infty} f_n \, h_n(t) ,\\
f_n  & = & \int dt w(t) \, f(t)\, h_n(t). 
\eea
The Hilbert space $H$ with the measure $dt\, w(t)$ is
spanned by the functions $\{ h_n(t) \}_{n=0}^{\infty}$.

The important consideration is that we take $w(t)$ to be a function
that is simultaneously a measure for an abstract Hilbert space 
and at the same time, 
$w(t)$ gives a good approximation
to the experimentally measured correlator $R_2(t)$, so that the
first few expansion coefficients would suffice to fully describe
the measured function. For example, $w(t)$ can be taken as a
(multi-dimensional) Gaussian function. In case of a one-dimensional
Gaussian function, the Hermite polynomials form a complete orthonormal
set of functions with respect to a Gaussian weight. Then we can
obtain a convergent expansion in terms of Gaussians and Hermite
polynomials as specified in Section \ref{s:edge}.
Note also that the dimensionless variable $t$ can be defined
in one, two or three dimensions, e.g. $t = R_I Q_I$ or
$t = (R_L Q_L, R_T Q_T)$ or $ t = (R_L Q_L, R_{side} Q_{side},
R_{out} Q_{out})$. The multi-dimensional parameterizations
will be addressed explicitly in Section~\ref{s:multi}.

Let us assume, that the function $g(t) = R_2(t)/w(t) $ is also an 
element of the Hilbert space $H$.
This is possible, if 
\be
	\int dt\, w(t) g^2(t) = \int dt\, \left[R_2^2(t) /w(t)\right] < \infty ,
		\label{e:gnorm}
\ee
i.e. if the experimentally measured correlator, $R_2(t)$ approaches
the value 0 sufficiently fast. With the help of a numerical integration
of the experimentally determined $R_2(t)$ function and the selected
$w(t)$ zero-th order approximation to $R_2(t)$, the convergence of the 
integral in the above eq.~(\ref{e:gnorm}) can be explicitly checked.

The function $g$ can be expanded as
\bea
	g(t) & = & \sum_{n =0}^\infty g_n h_n(t),\\
	g_n  & = & \int dt\, R_2(t) h_n(t) \label{e:numgn}. 
\eea
Thus the expansion coefficients $g_n$ can be determined by a simple
numerical evaluation of the integral in  eq. ~(\ref{e:numgn}),
using the experimentally determined values for the correlator $R_2(t)$.
 
From the completeness of the Hilbert space and from the assumption
that $g(t)$ is in this Hilbert space, i.e. eq. ~(\ref{e:gnorm})
is satisfied, 
one obtains the following series expansion for the correlator $R_2(t) $:

\be
	R_2(t) =  w(t) \sum_{n = 0}^\infty g_n h_n(t) .
\ee
It is obvious from this form, that the best choice of the 
$w(t)$ function is the form that is as close to the experimentally
determined correlator, as possible. 
For convenience, it is advantageous to choose $h_0(t) = const$
and use the normalization $g_0 h_0(t) = 1$, by an appropriate rescaling
of the magnitude of the function $w(t)$.
In such a case, the first few expansion coefficients may be sufficient
to characterize the correlation function and the correlator $R_2(t)$.

In order to characterize with an independent parameter the
strength of the two-particle correlator $R_2(t)$, the expansion-dependent
parameter $\lambda_w$ is introduced, and
for convenience, the convention $w(t = 0) = 1$ is 
utilized.  Thus the two-particle BECF can be
expanded into the following family of series:

\be
	C_2(t) = {\cal N}
	\left\{  
		1 +\lambda_w\,  w(t) \sum_{n = 0}^\infty g_n h_n(t)
	\right\} ,
	\label{e:cgene}
\ee
where the core-halo~\cite{chalo} intercept
parameter of the correlation function is
\be
	\lambda_* = \lambda_w \sum_{n = 0}^\infty g_n h_n(0) ,
\ee
and the coefficients of the expansion, $g_n$ can be determined
either from numerical evaluation of eq.~(\ref{e:numgn}) 
or from a fit to the measured data points with a particular 
concrete form of eq.~(\ref{e:cgene}).
	
The above general example can be utilized to study
correlation functions in various domains and with various
weight functions and related orthogonal polynomials,
for example the Legrende, associated Legrende polynomials,
spherical harmonics, Gegenbauer, Chebyshev, Hermite, Laguerre
or Jacobi polynomials.

In the next two Sections, we demonstrate the power of the
method by working out the specific examples for 
approximately exponential and approximately Gaussian correlators,
by using $w(t) = \exp(-t)$  in the $0 \le t < \infty$ domain and
and $w(t) = \exp(-t^2/2)$ in the $-\infty < t <  \infty $ domain.
As we develop a fitting method, we will utilize only the orthogonality
of the complete orthonormal set of functions 
$\left\{ h_n(t)\right\}_{n=0}^{\infty} $, 
giving a simple form for $H_n(t) \propto h_n(t)$ and
absorbing the normalization coefficients of $h_n(t)$ into the fit parameters.

\section{Laguerre expansion and exponential shapes
\label{s:lagu}}
If in a zeroth-order approximation the correlation function
has an exponential shape, then it is an efficient method to 
apply the Laguerre expansion, as a special case of
eq.~(\ref{e:cgene}):
\bea
	t & = & Q R_L, \\
	w(t)& = & \exp(-t),\\
	\int_0^{\infty} dt \, \exp(-t) L_n(t) L_m(t) &  \propto  & \delta_{n,m},
\eea
where the order-$n$ Laguerre polynomial is defined as
\be
	L_n(t) = \exp(t) \frac{d^n}{dt^n} (-t)^n \exp(-t).
\ee
The general form of eq.~(\ref{e:cgene}) takes the particular form
of the Laguerre expansion~\cite{edge-cst} as:
\bea
C_2(Q) = {\cal N} \left\{ 
	1 + \lambda_L \exp(- Q R_L) 
	\left[ 1 + c_1 L_1(Q R_L) + \frac{c_2}{2!} L_2(Q R_L) + ... \right]
	\right\}
	.  \label{e:laguerre}
	\nonumber \\
	&&
\eea
The fit parameters are the scale parameters ${\cal N}$,
$\lambda_L$, $R_L$ and the expansion coefficients $c_1$, $c_2$, 
\, ... \, .  
The first few Laguerre polynomials are explicitly given as
\bea
	L_0(t) & = & 1, \\
	L_1(t) & = & t  - 1,\\
	L_2(t) & = & t^2 - 4t + 2, \, ... \, .
\eea
The Laguerre polynomials are non-vanishing at the origin,
hence $C_2(Q = 0) \ne 1 + \lambda_L$.
The core-halo intercept parameter, $\lambda_*$,
is defined by the extrapolated value of the
two-particle correlation function at $Q = 0$, see 
refs.~\cite{chalo,cnhalo,dkiang,pcnhalo} for further details.
If the core/halo model is applicable, and effects of partial
coherence in the particle emitting source can be neglected,
the core-halo intercept parameter $\lambda_*$ measures the squared fraction
of bosons emitted directly from the core, 
an important physical observable that 
is related to the degree of partial restoration of $U_A(1)$ symmetry
in hot and dense hadronic matter~\cite{Csorgo:1999vd,Vance:1998wd},
as a partial $U_A(1)$ symmetry restoration  
leads to enhanced production of $\eta'$ mesons and suppression
of the core fraction of directly produced pions at low values of 
transverse momentum.

The physically significant core-halo intercept
parameter $\lambda_*$, and its error $\delta\lambda_*$ 
can be obtained from the 
parameter $\lambda_L$ of the Laguerre expansion as

\bea
	\lambda_* & = & \lambda_L [1 - c_1 + c_2 - ... ] , \\
	\delta^2 \lambda_* & = & 
	\delta^2 \lambda_L \left[ 1 + c_1^2 + c_2^2 + ... \right] +
	\lambda_L^2 \left[ \delta^2 c_1 + \delta^2 c_2 + ... \right] .
\eea
Any Bose-Einstein correlation function can be expanded into a 
{\it convergent} Laguerre expansion of the form of eq.~(\ref{e:laguerre}),
provided that
\be
	\int_0^\infty dt \, R_2^2(t) \exp(+t) < \infty,
\ee
where $t = Q R_L$ stands for the dimensionless scale variable. 
In principle, the left-hand side of this unequality can be evaluated
directly from the experimental data. 

\section{Edgeworth expansion and Gaussian shapes
\label{s:edge}}
If, in a zeroth-order approximation, the correlation function
has a Gaussian shape, then it is an efficient method to 
apply the Edgeworth expansion to characterize the deviation
from the Gaussian form in the following manner:

\bea
	t & = & \sqrt{2} Q R_E, \\
	w(t)& = & \exp(-t^2/2),\\
	\int_{-\infty}^{\infty} dt \, \exp(-t^2/2) H_n(t) H_m(t) &  
	\propto & \delta_{n,m},
\eea
where the order-$n$ Hermite polynomial is defined as
\be
	H_n(t) = \exp( t^2/2) \left( - \frac{d}{dt} \right)^n
		\exp(-t^2/2).
\ee
The general form of eq.~(\ref{e:cgene}) takes the particular form
of the Edgeworth expansion~\cite{edge0,edge,edge-cst} as:
\bea
C_2(Q) & = &{\cal N} \left\{ 
	1 + \lambda_E \exp( - Q^2 R_E^2) 
		\right. \times \nonumber \\
		&& 
	\left. 
	\left[ 1 + \frac{\kappa_3}{3!} H_3(\sqrt{2} Q R_E)
		+\frac{\kappa_4}{4!} H_4(\sqrt{2} Q R_E) + ... \right]
		\right\} .
		\label{e:edge}
\eea
The fit parameters are the scale parameters ${\cal N}$,
$\lambda_E$, $R_E$ and the expansion coefficients $\kappa_3$, $\kappa_4$, 
\, ... \, ,  that coincide with the cumulants of rank 3, 4, ...,
of the correlation function.
The first few Hermite polynomials are listed as
\bea
	H_1(t) & = & t, \\
	H_2(t) & = & t^2 -1, \\
	H_3(t) & = & t^3 - 3 t , \\
	H_4(t) & = & t^4 - 6 t^2 + 3,\, ... \,
\eea
The physically significant core-halo intercept
parameter $\lambda_*$ can be obtained from the 
Edgeworth fit of eq.~(\ref{e:edge}) as
\bea
\lambda_* & = & \lambda_E \left[ 1 + \frac{\kappa_4}{8} + ... \right],\\
\delta^2 \lambda_* & = & \delta^2 \lambda_E + 
	(\kappa_4 \delta\lambda_E + \lambda_E \delta\kappa_4)^2/64 + ... \, .
\eea
Any Bose-Einstein correlation function can be expanded into 
a {\it convergent} Edgeworth expansion, if
\be
\int_{-\infty}^\infty dt\, R_2^2(t) \exp(+ t^2/2)  < \infty .
\ee
This latter requirement can be checked experimentally, if necessary.
This expansion technique was applied  in the conference
contributions~\cite{edge,edge-cst} to the AFS minimum bias 
and 2-jet events to characterize
successfully the deviation of data from a Gaussian shape.
It was also successfully applied to characterize the 
non-Gaussian nature of the correlation function in two-dimensions
in case of the preliminary E802 data in ref.~\cite{edge-cst},
and it was recently applied to characterize the non-Gaussian nature
of the three-dimensional
two-pion BECF in $e^+ + e^-$ reactions at LEP - 1~\cite{L3}.

\section{Multi-dimensional expansions\label{s:multi}}
Genuine multi-dimensional expansions can be obtained by using multi-dimensional
complete orthogonal sets of functions. The simplest two-dimensional 
example is the study of angular correlations on the surface of the 
unit sphere (e.g. to study correlations of jets within jets in QCD).
In this case, $t = (\theta, \phi) $, the measure is $w(t) = \sin(\theta)$ 
and the domain is ($0 \le \theta \le \pi$, $0 \le \phi < 2 \pi$). 
The orthogonal polynomials are the well-known spherical harmonics
$Y_l^m(\theta,\phi)$ that satisfy
\begin{equation}
	\int_0^\pi d\theta \sin(\theta) \int_0^{2 \pi} d\phi  \, \,
	\overline{Y}_k^n(\theta,\phi) 
	Y_l^m(\theta,\phi)
	\propto \delta_{k,l} \delta_{m,n}
\end{equation}
and the angular correlations can be expanded in a series as
\begin{equation}
	C_2(\theta,\phi) = 1 + \lambda_Y \sin(\theta) [1 + \sum_{l = 1}^\infty
	\sum_{m = -l}^l c_{lm} Y_l^m(\theta,\phi) ].
\end{equation}
This example should be sufficient to demonstrate how the method can be
generalized to higher dimensions.
A systematic treatise of the multi-dimensional realizations
of the general expansion method as described in Section~\ref{s:gene}
is beyond the scope of the present paper. 

In the following we consider the special case of factorized multi-dimensional
distributions. 
A more profound technique can be obtained based on the specialization
of the general arguments of section 2 to the case of multi-dimensional
BECF-s. The two-dimensional Edgeworth expansion and the interpretation
of its parameters is described in the handbook on mathematical statistics
by Kendall and Stuart~\cite{kendallstuart}.

If the two-particle BECF can be factorized as a product of (two or more)
functions of one variable each, then e.g. a Laguerre or an Edgeworth
expansion can be applied to the multiplicative factors -- functions of
one variable, separately. This method was applied recently to
study the non-Gaussian features of multi-dimensional 
Bose-Einstein correlation functions in refs.
~\cite{edge,L3}.
 
 Let us restrict
 ourselves to the situation when the multi-dimensional 
 Fourier-transformed emission function factorizes in certain set of variables.
 For example, such a boost-invariant form of factorization is
 assumed in the Buda-Lund parameterization of the two-particle
 Bose-Einstein correlation functions in refs.~\cite{bl,cf98-bl}.

A simple expression for the two-particle correlation functions is
obtained for longitudinally approximately boost-invariant,
cylindrically symmetric particle emitting sources,
in the \BL~picture, if we assume that  
the emission function factorizes
as a product of 
an effective proper-time distribution, a space-time rapidity
distribution and a transverse coordinate distribution
 where $\eta = 0.5 \log[ (t + r_z) / (t - r_z)]$ is the space-time
 rapidity, $\tau = \sqrt{t^2 - r_z^2} $ is the longitudinally
  boost-invariant proper-time distribution~\cite{bl,cf98-bl}.

	With the help of a small source size 
	(or large relative momentum)
	expansion, the two-particle Bose-Einstein correlation function
	can be written into the following boost-invariant,
	factorized \BL~form: 
\bea
	C_2(\Delta k, K) & = & 1 + \lambda_*(K) 
		{\dst |\tilde H_*(Q_=)  |^2 \ov |\tilde H_*(0) |^2} \,
		{\dst |\tilde G_*(Q_{\parallel})  |^2 \ov |\tilde G_*(0) |^2}\,
		{\dst |\tilde I_*(Q_{\perp})|^2 \ov 
		|\tilde I_*(0) |^2} .
	\label{e:blf}
\eea
	Here, the Fourier-transformed proper-time	,
	space-time rapidity and transversal coordinate
	distributions, $\tilde H_*(Q_=) $,
	$\tilde G_*(Q_{\parallel}) $ and
	$\tilde I_*(Q_{\perp})$
	can be of power-law, exponential,
	Gaussian or other types, corresponding to the
	underlying space-time structure of the particle
	emitting source, 
	hence it is natural to apply the one-dimensional 
	Edgeworth or the Laguerre
	expansions to any of these factors. 

	The invariant {\it temporal}, {\it parallel} and
	{\it perp}endicular 
	relative momentum components are defined, respectively, as  
\bea
	Q_= & = &   
		Q_0 \cosh[\etab] - Q_z \sinh[\etab]\, \equiv  \,Q \cdot \nb ,
		 \label{e:q=} \\
	Q_{\parallel} & = & 
		 Q_0 \sinh[\etab] - Q_z \cosh[\etab]  
		\, \equiv \,
		 Q \times \nb , \label{e:qpar} \\
	Q_{\perp} & = & \sqrt{ Q_{x}^2 + Q_{y}^2},   	\\
 	Q^2 & = & Q \cdot Q \, = \,  
	(Q_=)^2 - (Q_\parallel)^2 - Q_{\perp}^2. \,   \label{e:qperp}
\enn
	In the above, 
	$a \cdot b = a^{\mu} b_{\mu} = a_0 b_0 - {\bf a} {\bf b}
	 = a_0 b_0 - a_x b_x - a_y b_y - a_z b_z$
	stands for the inner product of four-vectors.
	The direction of the normal-vector
	$\nb$ is characterized as 
	$\nb^{\mu} = (\cosh[\etab],0,0,\sinh[\etab])$, by a mean space-time 
	rapidity $\etab$~\cite{bl,cf98-bl} in the LCMS frame~\cite{lcms}. 
	Hence this direction is boost-invariant and $\etab$ 
	is one of the fit parameters.	
	As $\nb \cdot \nb = +1$, this normal-vector points into a
	time-like direction and $Q \cdot \nb =  Q_=  $ is an 
	invariant time-like component of the relative momentum.
	The relative momentum component	$Q_{\parallel}$ is 
	parallel to the longitudinal direction 
	(which is the thrust axis in jet physics, or the beam
	direction in heavy ion physics),
	and $Q_{\parallel}$ is invariant to boosts along this direction.
	Finally, $Q_{\perp}$ is the remaining perpendicular (or transverse )
	component of the relative momentum, it is also invariant for
	longitudinal boosts.

	In eq.~(\ref{e:blf}) the Fourier-transformed distributions
	are defined and explained in greater detail 
	in refs.~\cite{bl,cf98-bl}. 
	Although these distributions can be theoretically evaluated
	from an assumed shape of the emission function
	$S(x,k)$, for example see ref~\cite{bl,cf98-bl},
	here we are interested in the model-independent characterization
	of the Bose-Einstein correlation functions, exclusively.

	From a three-dimensional analysis of two-pion Bose-Einstein
	correlation data at LEP1 ~\cite{L3} we know that
	the BECF is approximately a factorized Gaussian.
	Thus, a good candidate to characterize the 
	non-Gaussian nature of these correlations in
	3 dimensions, in a boost-invariant manner, is
	the following Edgeworth expansion
	of the factorized form of the \BL correlation function:
\bea
	C_2(\Delta k, K) & = & 1 + \lambda_E
	\exp( - Q_=^2 R_=^2 - Q_{||}^2 R_{||}^2 - Q_{\perp}^2 R_{\perp}^2
		) \times \nonumber\\
	&& \left[ 1 + \frac{\kappa_{3,=}}{3!} H_3(\sqrt{2} Q_{=}R_{=} )
	+ \frac{\kappa_{4,=}}{4!} H_4(\sqrt{2} Q_{=}R_{=} ) + ... \right]
			\times\nonumber \\
	&& \left[ 1 + \frac{\kappa_{3,||}}{3!} H_3(\sqrt{2} Q_{||}R_{||} )
	+ \frac{\kappa_{4,||}}{4!} H_4(\sqrt{2} Q_{||}R_{||} ) + ... \right]
			\times\nonumber \\
	&& \left[ 1 + \frac{\kappa_{3,\perp}}{3!} 
			H_3(\sqrt{2} Q_{\perp}R_{\perp} )
	+ \frac{\kappa_{4,\perp}}{4!} 
	H_4(\sqrt{2} Q_{\perp}R_{\perp} ) + ... \right] .
	\label{e:bl-edg}
\eea

	What are the fit parameters in eq.~(\ref{e:bl-edg})?
	We have {\it $5$ free scale parameters} for cylindrically
	symmetric, longitudinally expanding sources,
	and {\it three series of shape parameters}.
	The scale parameters are
	$\lambda_E$, $R_=$, $R_{\parallel}$, $R_{\perp}$ and $\etab$,
	that characterize the effective source at a given mean momentum,
	by giving the vertical scale of the correlations, the invariant
	temporal, longitudinal and transverse extensions of the source
	and its invariant direction, which is
	the space-time rapidity of the effective source in the LCMS frame
	(the frame where $k_{1,z} + k_{2,z} = 0$, ~\cite{lcms}). 
	The three series of shape parameters are
	$\kappa_{3,=}$, $\kappa_{4,=}$, ... ,
	$\kappa_{3,||}$, $\kappa_{4,||}$, ... ,
	$\kappa_{3,\perp}$, $\kappa_{4,\perp}$, ... .

	In principle, each of these parameters may depend on the mean momentum
	$\bK$. At any fixed value of the mean momentum $\bK$,
	the above  free parameters of the invariant \BL correlation function
	can be fitted to data, without any a priori assumption on the
	shape of the source emission function, based  only on the
	cylindrical symmetry and the boost-invariant factorization
	property of the particle emitting source.

\section{Application of the results}
	As the Laguerre and the Edgeworth expansions of Bose-Einstein
	correlation functions were introduced with quite some success
	in refs.~\cite{edge,edge-cst} to characterize the AFS 2-jet and the
	AFS minimum-bias data in 1 dimension, to characterize the non-Gaussian
	features of the E802 preliminary $Si + Au \rightarrow \pi^+ + \pi^+
	+ X $ data in two dimensions~\cite{edge-cst}, and they were reported to 
	successfully characterize the non-Gaussian
	features of the L3 two-pion BECF at LEP1 in two and three dimensions,
	~\cite{L3} (although using a non-invariant formulation),
	we think that there is no need to further demonstrate the
	applicability of the Edgeworth expansion and its multi-dimensional
	generalizations in correlation studies in high energy physics.

	However, the applications of the Laguerre expansions were not yet
	presented in the literature as far as we know.
	Hence, let us demonstrate in Figure 1 the power of the
	Laguerre expansions to characterize two well-known,
	non-Gaussian correlation functions: the $C_2(Q_I)$ 
	correlation functions as determined 
	by the UA1 and the NA22 experiments~\cite{na22,ua1}.
	Note that the invariant momentum difference
	$Q^2$ is binned in logarithmic scale, and the resolution of both
	experiments goes down to about 30 MeV, a very small scale as
	compared to the typical 200 MeV half-width of the BECF in
	particle physics.

We note, that Fig. 1. is just an illustration of the power
of Laguerre expansions, however, the authors do not intend to
replace the power-law description of these correlation functions
by the Laguerre expansion, 
and the authors are aware of the fact that a power-law fit
gives a similarly good $\chi^2/NDF$ with smaller number of fit parameters.

\begin{table}[hbt]
\newlength{\digitwidth} \settowidth{\digitwidth}{\rm 0}
\catcode`?=\active \def?{\kern\digitwidth}
\caption{Best fits to UA1 and NA22 two-particle correlations 
using a Laguerre expansion}
\label{tab:results}
\begin{center}
\begin{tabular*}{\textwidth}{@{}|l@{\extracolsep{\fill}}|rl|rl|}
\hline
                 & \multicolumn{2}{l|}{UA1} 
                 & \multicolumn{2}{l|}{NA22} \\
\cline{2-5}  
		 Parameter~\,~\,~\,
                 & \multicolumn{1}{r}{Value} 
                 & \multicolumn{1}{l|}{Error} 
                 & \multicolumn{1}{r}{Value} 
                 & \multicolumn{1}{l|}{Error} 
		 \\
\hline
${\cal N}$         & 1.355 &$\pm$ 0.003    & 0.95  &$\pm$ 0.01      	\\
$\lambda_{L}$ & 1.23  &$\pm$ 0.07     & 1.37  &$\pm$ 0.10 		\\
$R_L$ [fm]  & 2.44  &$\pm$ 0.12     & 1.35  &$\pm$ 0.14		\\
$c_1$ 		   & 0.52  &$\pm$ 0.03     & 0.63  &$\pm$ 0.06  	\\
$c_2$ 		   & 0.45  &$\pm$ 0.04     & 0.44  &$\pm$ 0.06  	\\
\hline
$\chi^2/NDF$     & \multicolumn{2}{l|}{41.2/41 = 1.01} 
                 & \multicolumn{2}{l|}{20.0/34 = 0.59}  \\
\hline
\end{tabular*}
\end{center}
\end{table}

From Table 1, one can determine that the quality of the
Laguerre expansion fits is statistically acceptable both
in case of the UA1 and in case of the NA22 data. We can also
determine the core-halo model intercept parameter,
$\lambda_* = 1.14 \pm 0.10$ (UA1) and $\lambda_* = 1.11 \pm 0.17$
(NA22). As both of these values are within errors equal to unity,
the maximum of the possible value of the intercept parameter
$\lambda_*$ in a fully chaotic source, we conclude that
either there are other than Bose-Einstein short-range correlations
observed by both collaborations, or in case of this measurement
the full halo of long-lived resonances is resolved~\cite{chalo,dkiang}.
This is very interesting and the distribution of long-lived resonance
fractions and decay times may be related to the approximate
power-law shape of the fitted correlation functions~\cite{bialas}.
The full resolution of the halo is theoretically possible
only because the $\eta'$-s and $\eta$-s were either produced
in a negligible number, or because their effects were corrected
for in the Monte-Carlo evaluation of the background distribution.

\section{Summary}
In this paper we described a general method to characterize, in a 
model and theory independent manner the shape of the 
two-particle Bose-Einstein correlation functions with the help of  
mutually orthogonal sets of functions, assuming that 
the  measure of the abstract Hilbert space can be
identified with the approximate shape of the correlation
function. For one-dimensional, approximately Gaussian 
correlation data, the Edgeworth expansion technique is obtained.
For one-dimensional, approximately exponential shapes,
we obtained the expansion in terms of the Laguerre polynomials.
We generalized these expansion techniques to two and three dimensions.
We also demonstrated that the method works well:
we characterized the deviation of the two-particle Bose-Einstein
correlation function of the UA1 and the NA22 collaborations  from the
exponential shape with the first two Laguerre coefficients.

\section*{Acknowledgments}
Cs. T. is greatful to professor W. Kittel and the L3 collaboration
for an inspiration to complete this paper. 
This research has been supported by an NWO - OTKA grant N25487,
by the Hungarian NSF grants OTKA T024094 and T026435.

\vfill
\begin{figure}[ht]
\centering
\epsfig{file=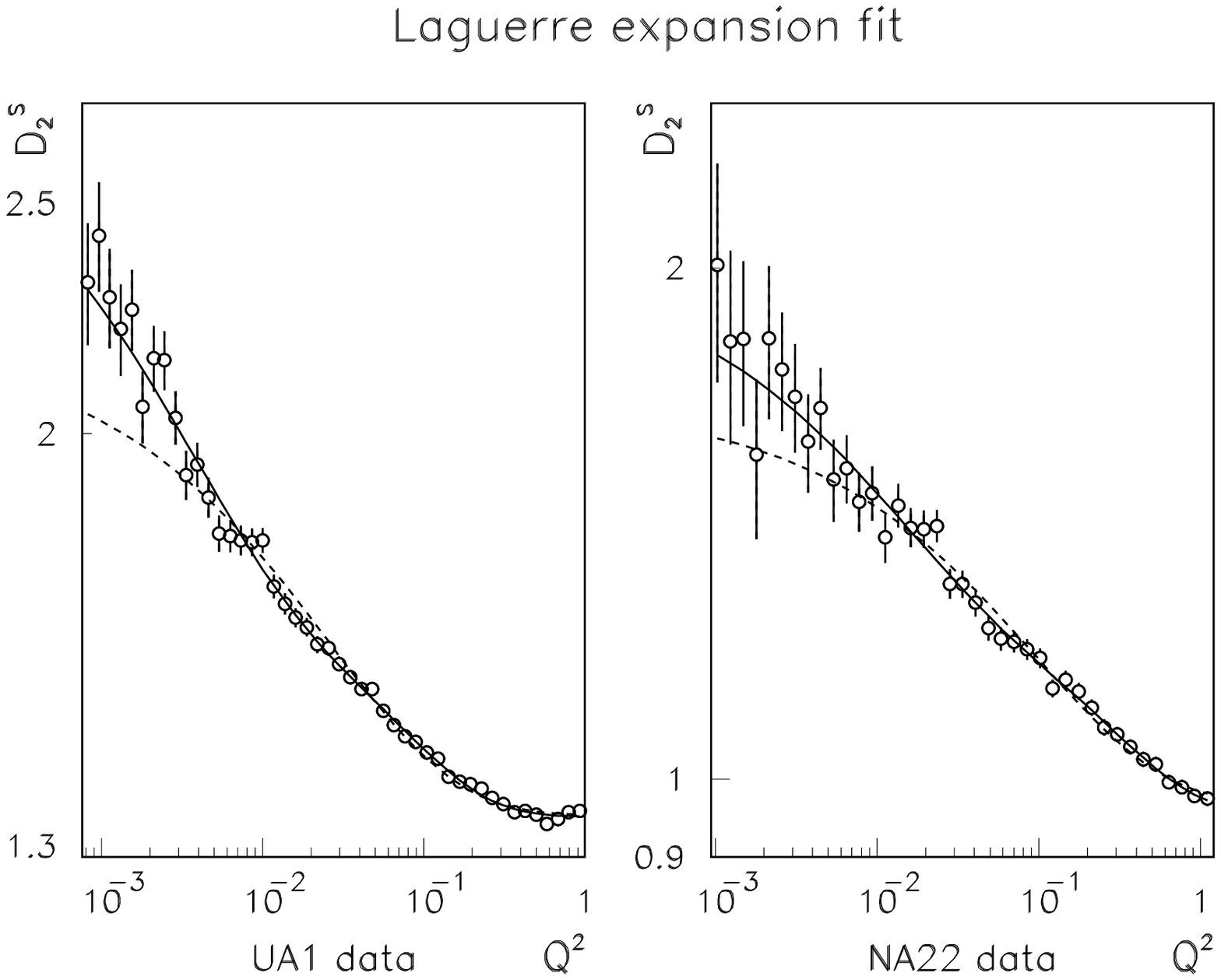,height=16cm}
\label{f:lagu}
\vspace{-2cm}
\caption{The figures show $D_2^s$ which is proportional to
the two-particle Bose-Einstein correlation
function, as measured by the UA1 and the NA22 collaborations.
The dashed lines stand for the exponential fit,
which clearly underestimates the measured points at low value of
the squared invariant momentum difference $Q^2_I$ (note the logarithmic
horizontal scale).
The solid lines stand for the fits with the Laguerre expansion method,
which is able to reproduce the data with a statistically
acceptable $\chi^2/NDF$. The fit results are summarized in Table 1.
}
\end{figure}
\vfill\eject 
\end{document}